
%
%
\input harvmac.tex
\def\nl{\hfill\break}
\def\<{\langle}
\def\>{\rangle}

\nref\ZamoI{A.B. Zamolodchikov, JETP Lett. 43 (1986) 730.}
\nref\CarLud{A. Ludwig, J. Cardy, Nucl.  Phys. B285 (1987) 687.}
\nref\Zamo{Al.B. Zamolodchikov, Nucl.  Phys.  B358 (1991) 497, 524.}
\nref\DF{V. Dotsenko, V.A. Fateev, Nucl. Phys. B240 (1984) 312.}
\nref\EY{T. Eguchi, S.K. Yang, Phys. Lett. 224B (1989) 373.}
\nref\HM{T. Hollowood, P. Mansfield, Phys. Lett. 226B (1989) 73.}
\nref\Nien{B. Nienhuis, J. Stat. Phys. 34 (1984) 153.}
\nref\F{G. Felder, Nucl.  Phys. B317 (1989) 215.}
\nref\GS{C. Gomez, G. Sierra, Phys.  Lett. B240 (1990) 149.}
\nref\BP{P. Bouwknegt, J. McCarthy, K. Pilch, Nucl. Phys. B352 (1991) 139.}
\nref\EYI{T. Eguchi, S.K. Yang, in Proceedings of the eighth symposium
on theoretical physics, Kyohak Yunkusa, Ed. H. S. Song (1989).}
\nref\Sm{N.Yu. Reshetikhin, F.A. Smirnov, Comm. Math. Phys. 131 (1990) 157.}
\nref\Lecl{A. LeClair, Phys. Lett. B230 (1989) 103.}
\nref\BL{D. Bernard, A. LeClair, Comm. Math. Phys. 142 (1991) 99.}
\nref\FL{G. Felder, A. LeClair, Int. J. Mod. Phys. A7 (1992) 239.}
\nref\P{V. Pasquier, Comm. Math. Phys. 118 (1988) 355.}
\nref\PS{V. Pasquier, H. Saleur, Nucl. Phys. B330 (1990) 523.}
\nref\Zamopol{A.B. Zamolodchikov, Mod. Phys. Lett. A6 (1991) 1807.}
\nref\Smir{F.A. Smirnov, Phys. Lett. B275 (1992) 109.}

\Title{\vbox{\baselineskip12pt\hbox{BUHEP-93-5, USC-93/003, LPM-93-07}
        }} {\vbox{\centerline{Massless Flows I:}
\vskip4pt\centerline{the sine-Gordon and $O(n)$ models}}}
\centerline{P. Fendley$^\#$, H. Saleur$^\dagger$ and Al.B. Zamolodchikov$^*$}
\smallskip\bigskip\centerline{$^\#$Department of Physics, Boston University}
\centerline{590 Commonwealth Avenue, Boston, MA 02215, USA}
\bigskip\centerline{$^\dagger$Department of Physics,
University of Southern California}
\centerline{Los Angeles, CA 90089, USA}
\bigskip
\centerline{$^*$Laboratoire de Physique Math\'ematique}
\centerline{Universit\'e Montpellier 2, Montpellier, France}
\vskip .14in
\baselineskip=14pt
The massless flow between successive minimal models of conformal field theory
is related to a flow within the sine-Gordon model when the coefficient of the
cosine potential is imaginary.  This flow is studied, partly numerically, from
three different points of view. First we work out the expansion close to the
Kosterlitz-Thouless point, and obtain roaming behavior, with the central
charge going up and down in between the UV and IR values of $c=1$. Next we
analytically continue the Casimir energy of the massive flow (i.e.\ with real
cosine term). Finally we consider the lattice regularization provided by the
$O(n)$ model in which massive and massless flows correspond to high- and
low-temperature phases.

A detailed discussion of the case $n=0$ is then given using the underlying
$N$=2 supersymmetry, which is spontaneously broken in the low-temperature
phase. The ``index'' $\tr\, F(-1)^F$ follows from the Painlev\'e III
differential equation, and is shown to have simple poles in this phase. These
poles are interpreted as occuring from level crossing (one-dimensional phase
transitions for polymers).  As an application, new exact results for the
connectivity constants of polymer graphs on cylinders are obtained.  These
results and points of view are used in the following paper to discuss the
appropriate exact $S$-matrices and the resulting Casimir energies.

\smallskip
\Date{April 1993}
\vfill\eject
\baselineskip=16pt plus 2pt minus 1pt
\newsec{Introduction}

Prompted by conformal field theory, there has been enormous progress in the
last several years in understanding two-dimensional field theories off the
critical point. For both massive and massless models, much has been understood
by using perturbed conformal field theory. In the numerous cases where the
model is integrable, a number of non-perturbative techniques have been
extensively developed and applied.  There are many massive models whose
properties are very well understood, but there exist some very basic massless
ones whose properties have not been fully explored.

A particular example is the flow between the $(t+1)^{\hbox{th}}$ and
$t^{\hbox{th}}$ unitary models \ZamoI, which has been studied perturbatively
for large $t$ \CarLud\ and for finite $t$ using the conjectured thermodynamic
Bethe ansatz relations \Zamo.  Moreover, this problem is related to an even
less-understood model, which is the sine-Gordon model with an imaginary
potential. This model is interesting in its own right, because at a
particular coupling it describes the ``dense'' phase of self-avoiding
polymers. These models are related because here and in many other cases, one
can consider perturbed conformal field theories as restrictions of affine Toda
theories, thus extending free-field representations of conformally invariant
field theories.  In the conformal case and the massive case, these
restrictions are well known.  The (Virasoro) minimal conformal field theories
can obtained from a free boson \DF\ and the massive integrable perturbation of
these theories by the $\Phi_{13}$ operator can be obtained from sine-Gordon
\refs{\EY,\HM}. In statistical-mechanical language, this is often called
mapping the system onto a Coulomb gas \Nien.  The restriction is most easily
understood in field theory using quantum-group structures \refs{\F -\FL}, and
to some extent parallels lattice-model constructions \refs{\P,\PS}. A very
similar restriction applies also to the massive $O(n)$ model
\refs{\Zamopol,\Smir}.

We discuss a similar truncation for massless flows, where both the UV and IR
limits are conformally invariant. In this case, since both UV and IR fixed
points are obtained by restriction of a free boson, one expects that the full
trajectory is the restriction of an intriguing flow in sine-Gordon, with
central charge equal to 1 at both extremities. This flow actually was first
mentioned in \Nien, and related explicit field-theory calculations were done in
\ref\Gri{M.T. Grisaru, A.  Lerda, S.  Penati and D. Zanon, Phys. Lett. B234
(1990) 88; Nucl. Phys. B342 (1990) 564.}.  We discussed it briefly in
\ref\FS{P. Fendley, H. Saleur, Nucl. Phys. B388 (1992) 609.} where we presented
numerical data with a ``roaming'' trajectory that violates the
$c_{\hbox{eff}}$ conjecture. We describe in this paper and in the following
our attempts to understand quantitatively this problem.

The models describing the flows between the minimal models are unitary, but
the sine-Gordon model with imaginary potential has a Lagrangian that is not
real, and hence is non-unitary. The non-unitarity allows a variety of
interesting properties to appear, including level crossing, an increasing
$c$-function, spontaneous breaking of supersymmetry, and singularities. While
these features are part of what makes this model interesting, they also often
make it difficult. In this paper and its sequel we will use a variety of
methods to explore the situation. We believe that we have completed the
picture for the minimal models; we will give the exact $S$-matrices and
ground-state energy. However, we will see that there are still a number of
unanswered questions in the case of sine-Gordon.  The first paper contains
detailed studies (including numerical) of the problem, excepting attempts to
build a scattering theory. The second paper is then fully devoted to that
question.

In section 2 we discuss the flow within sine-Gordon in an expansion
close to the Kosterlitz-Thouless point. We recover the roaming trajectory with
a pair of extrema of order $1/t^3$, and discuss the behavior of the
stress-energy tensor and alternate definitions of the central charge.

In section 3 we discuss in detail the lattice regularization of the problem
provided by the $O(n)$ model on the honeycomb lattice flowing to the
low temperature phase. A quantum-group symmetry present all along the
flow is exhibited that will provide useful information about level
crossings. A lattice derivation of the flow of independent sectors of the
partition functions is also provided. This section may be skipped by the
reader who is not interested in lattice questions.

In section 4 we discuss numerically the massless flow using two approaches.
We discuss first the analytic continuation of the massive flow \FS.  The first
extremum of the central charge agrees nicely with the expansion around the
Kosterlitz-Thouless point, but breaks down afterward. We also consider lattice
data, which show the existence of multiple extrema for small enough $t$, a
feature not seen in the large-$t$ expansion.

Section 5 contains a short discussion of the boundary case $n=-2$, which is
exactly solvable by other methods. The behavior is somewhat pathological, but
the appearance of singularities provides compelling evidence for similar
behavior to occur at least at nearby values of $n$.

\nref\CV{S. Cecotti, C.Vafa, Nucl.  Phys. B367 (1991) 359.}
\nref\CVFI{S. Cecotti, P. Fendley, K. Intriligator, C. Vafa, Nucl. Phys. B386
(1992) 405.} Section 6 contains a detailed discussion of the case $n=0$. This
case is physically very important because it describes polymers, and is also
quite amenable by analytic techniques due to the underlying $N$=2
supersymmetry. We find that $N$=2 supersymmetry is spontaneously broken; the
theorem that ordinarily does not allow this does not apply as a result of
non-unitarity (despite the non-unitarity, the described model is meaningful
for condensed-matter physics). It is shown that an infinite number of level
crossings separate the UV and IR fixed points. These crossings are studied
analytically by calculating the index $\tr\, F(-1)^F$ \refs{\CV,\CVFI}, where
they show up as poles. The locations of these poles are found by solving the
Painlev\'e III differential equation in the complex plane.

\newsec{Thermal flow: generalities and the large-$t$ expansion}

Consider the $(t-1)^{\hbox{th}}$ minimal unitary conformal field theory with
central charge \ref\BPZ{A.A. Belavin, A. Polyakov, A.B. Zamolodchikov, Nucl.
Phys. B241 (1984) 333.}
\eqn\cmin{c=1-{6\over t(t+1)},}
where $t$ is an integer.  As is well known, it can be obtained by twisting and
reducing a Gaussian model \DF\ (a free scalar field $\phi$) which has coupling
in ``lattice-like'' conventions $g=1+{1\over t}$ \Nien.

We perturb this critical theory by the thermal operator, which is identified
in the conformal language as $\Phi_{13}$, and has left and right dimensions
\eqn\hterm{h=\bar h={t-1\over t+1}.}
Perturbation by $\Phi_{13}$ corresponds in the Gaussian model to a
perturbation by the operator $\hbox{cos }2\phi$ with weight, with respect to
the untwisted stress energy tensor
\eqn\htermI{h=\bar h={t\over t+1}.}
The coupling constants are different due to the difference in
conformal weights in the twisted and untwisted theory. If for the minimal
model
\eqn\onpert{S_{\hbox{min}}=S^*_{\hbox{min}}+(\delta\beta)\int \Phi_{13}\
d^2 x,}
then
\eqn\sgpert{S_{SG}=\int d^2 x \left[{g\over 4\pi}(\del\phi)^2 +
 (\delta\beta)^{1/2}\hbox{cos }2\phi\ \right] .}
The mass in the minimal model is proportional to $|\delta\beta|^{{t+1\over
4}}$, where $\delta\beta=\beta_c-\beta$ refers to distance from the critical
inverse ``temperature''.

For $\delta\beta>0$ the minimal model is expected to flow to a trivial massive
phase. From the Gaussian point of view, we indeed have the standard
sine-Gordon model.  It is integrable, with a known $S$-matrix, and only the
massive soliton and antisoliton are in the spectrum for $t\geq 2$
\ref\ZandZ{A.B. Zamolodchikov and Al.B. Zamolodchikov, Ann.  Phys. 120 (1980)
253}.  {}From this $S$-matrix and the thermodynamic Bethe ansatz (TBA),
relations can be written for the ground-state energy which restrict to the
ones conjectured for the minimal models \Zamo\ when appropriately twisted \FS.

When $\delta\beta<0$ the minimal model flows \refs{\ZamoI,\CarLud}\ to another
conformal field theory with central charge in the infrared
\eqn\newc{c=1-{6\over t(t-1)}}
The corresponding situation in the sine-Gordon theory is a little unusual.  It
leads in \sgpert\ to a purely imaginary coupling, a non-unitary case that has
already attracted some attention \nref\Boy{D. Boyanovsky, R. Holman, Nucl.
Phys. B332 (1990) 641.}\refs{\Nien,\Gri,\Boy}.  The behavior can be understood
qualitatively as follows.  Call $\lambda_e$ the coupling of the
vertex operator $e^{ie\phi}$ in a general action with the Gaussian term plus
perturbations. In \sgpert\ for instance we have
$\lambda_2=\lambda_{-2}={1\over 2}(\delta\beta)^{1/2}$. Approximate
renormalization group equations can be written for the $\lambda_e$ and the
coupling constant $g$\nref\Jose{J.V. Jose, L.P. Kadanoff, S. Kirkpatrick,
D.R. Nelson, Phys. Rev. B16 (1977) 1217.}
\nref\Car{J. Cardy, in ``Fields, Strings and Critical Phenomena'', Les Houches
Proceedings (1988), ed. by E. Brezin and J. Zinn-Justin, North Holland.}
\refs{\Jose,\Car}.  If $b$ is the rescaling parameter, one finds at first
order
\eqn\renl{{d\over db}\lambda_e=\left(2-{e^2\over 2g}\right)\lambda_e,}
and for the coupling constant
\eqn\reng{{1\over g}{d\over db}g= 8\pi^2\lambda_e\lambda_{-e}.}
No magnetic charges arise in these equations.  Observe that $\hbox{cos }2\phi$
is relevant for $g>1$, irrelevant otherwise.

In the case $\delta\beta>0 $, $\lambda_2=\lambda_{-2}$ are real numbers. Then
if initially $g>1$ as in the case here, $\lambda_2$ and $\lambda_{-2}$ both
increase along the renormalization group flow \renl. This increase contributes
to increasing $g$ as well due to \reng\ and in turn to making $\hbox{cos
}2\phi$ even more relevant.  So at this order, more and more charges $\pm 2$
are generated at large distance, and one expects (slightly changing the
language) the ``Coulomb gas'' \refs{\Nien,\Jose}\ to flow to a massive, plasma
phase.

Now suppose $\delta\beta<0$. Initially, the magnitudes of both $\lambda_2$ and
$\lambda_{-2}$ both increase in the renormalization.  However, since they are
purely imaginary, their product is negative, so due to \reng\ the coupling
decreases. This decrease means that the operator $\hbox{cos }2\phi$ becomes
less relevant. One therefore expects that floating charges $\pm 2$ are present
in large numbers at intermediate distances, but ultimately disappear in the
IR. If this is so, the sine-Gordon model with imaginary coupling should flow
back to a Gaussian model, although with a smaller coupling constant. This
phenomena was first discussed in \Nien\ in relation to the dense $O(n)$ model
to be discussed in subsequent sections.

There is a simple way of understanding this heuristically \refs{\EY,\HM,\Gri}.
This is to postulate that the action for the conformal minimal model in
\onpert\ is in fact the Liouville action with a curvature term to represent
the charge at infinity:
\eqn\liouville{S^*_{\hbox{min}}=\int  {g\over 4\pi}\left[(\del\phi)^2 +
iQ R \phi \right] + \lambda_{2} e^{2i\phi} ,}
where $R$ is the scalar curvature. This model is reasonably well-understood
(at least with $Q$ imaginary, in our conventions) and has a fixed point which
is a conformal theory of central charge $c=1-6 g Q^2$. Thus the conjecture is
that with appropriate choices of $g$ and $Q$ (real here), this action
describes a minimal model. The operator with dimensions of $\Phi_{1,3}$ turns
out to be $\exp(-2i\phi)$. Thus to perturb the model, we add $\lambda_{-2} \int
\exp(-2i\phi)$ to the action, and obtain the sine-Gordon model \sgpert\
plus a curvature term. Thus in this sense, the actions \sgpert\ and \onpert\
differ only by the curvature term. Calculations in field theory including the
curvature term have been done for large $t$ in \Gri; these indeed reproduce
the flow of the central charge between successive minimal models.  The sign of
the perturbation is crucial: we can always redefine the coefficients of the
exponential terms by shifting the field $\phi$, but the product $\lambda_2
\lambda_{-2}$ does not change. Thus if this product is positive we are in the
massive phase, while if it is negative we are in the massless phase.

\subsec{Large-$t$ expansion}
It is possible to study this a little more quantitatively in the limit of
large $t$. Setting $e=2$ in \renl\ and \reng, both the fugacities and the
renormalization group eigenvalue $y_e=2-{e^2\over 2g}$ are small in this
limit, and we can simultaneously expand in both variables. At dominant order
these two equations allow us to compute
\eqn\dgdgdb{{d\over dg}{dg\over db}=4(g-1),}
which we can integrate to
\eqn\intdg{{dg\over db}=2\left[(g-1)^2-{1\over t^2}\right]}
where the integration constant is determined by the UV value $g_{UV}=1+{1\over
t}$. This beta function vanishes at another value of $g$, indeed smaller than
one as expected, $g_{IR}=1-{1\over t}$ (this value is in fact exact, but all
we get here is a check of it at first order).  The flow can be
integrated to give
\eqn\bb{g={g_{UV}+g_{IR}e^{4(b-b_0)/t}\over 1+e^{4(b-b_0)/t},}}
and
\eqn\lamb{\lambda_2=-\lambda_{-2}={i\over\pi t}{e^{2(b-b_0)/t}\over
1+e^{4(b-b_0)/t}.}}
Both the running coupling constant $g$ and the fugacities exhibit the
behavior qualitatively described in \Nien. In particular there is a value of
$b=b_0$ where the fugacities are extremal, so that their associated beta
functions vanish. However, this is not an intermediate fixed point at this
order, because the beta function for the coupling constant $g$ is not zero but
instead extremal while $g$ passes through $g=1$.

Further information about this flow in the Gaussian model is obtained by
considering the central charge. Recall that at first order the trace of the
stress-energy tensor for the Gaussian model perturbed by vertex operators of
charge $\pm e$ reads \Car
\eqn\trace{\Theta=-2\pi y_e\left(\lambda_e e^{ie\phi}+\lambda_{-e}
e^{-ie\phi}\right).}
{}From this and the relation between the derivative of the running central
charge defined in \ZamoI\
and $\Theta$ we find the running central charge at distance $R$
\eqn\cR{c(R)=1-6\pi^2y_e\left(R^{y_e}\lambda_eR^{y_e}\lambda_{-e}\right),}
Put now $e=2$.  When $\delta\beta>0$ the product of fugacities is positive, so
$c$ decreases in agreement with the $c$-theorem. On the other hand, when
$\delta\beta<0$ the $c$-theorem is not expected to hold due to the
non-unitarity of the theory.\foot{Here this non-unitarity manifests itself
rather trivially by the negative sign of the two point function of $\Theta$.}
In fact, one has the opposite behavior; since the couplings are purely
imaginary, $c$ initially increases above its UV value 1, going against a
``$c_{\hbox{eff}}$-theorem'' (this was observed in \Boy). Ultimately, $c$ must
return to 1 in the IR.  At the order to which we are working, $c$ is given by
\eqn\clart{c=1+{12\over t^3}{e^{4(b-b_0)/t}(1-e^{4(b-b_0)/t})\over \left(
1+e^{4(b-b_0)/t}\right)^3},}
and it exhibits a pair of extrema at $e^{4(b-b_0)/t}=(2 \pm\sqrt{3})$
with values
\eqn\cext{c_{\pm}=1\pm {2\over \sqrt{3}}{1\over t^3}.}
The $R$ dependence disappears at this order.

As a result of the vanishing of the beta functions for the fugacities, the
two-point functions $\<\Theta\Theta\>$ and $\<T\Theta\>$ vanish when $g$ passes
through $g=1$ at this order. It is likely that higher-order contributions give
them a finite but small value.  Indeed, assuming that the coupling $g$ evolves
monotonically, its beta function never vanishes, and there cannot be any
intermediate fixed point.

Since $c(R)$ is equal to one both in the UV and IR where the sine-Gordon model
flows to a Gaussian model, for any value of $R$ it must exhibit at least a
pair of extrema. Using the Callan-Symanzik equation
\eqn\CalZ{R{\partial c\over\partial R}+\beta_{g_i}{\partial c\over \partial
g_{i}}=0}
for coupling constants $g_i$ (here $g,\lambda$) we see, without any reference
to a given order of perturbation, that at these extrema the derivative
${\partial c\over \partial R}$ has to vanish, which in turn means vanishing of
$\<\Theta(0)\Theta(R)\>$. However we can expect a priori that the peculiar
values of the couplings $g_i$ at the extrema depends on $R$ so there is no
special intermediate point in the flow where $\<\Theta(0)\Theta(R) \>$ would
vanish for every $R$.  It is possible for such a two-point function to vanish
for some value of $R$ because of the non-unitarity of the theory.

Beyond this order, the perturbative study of the flow looks quite complicated.
However, according to the arguments of \Zamo, this model is integrable, and
non-perturbative methods should be applicable. Recall indeed that in the
massive flow, the running central charge $c(mR)$ ($m$ is the soliton
mass) for the sine-Gordon model was successfully calculated by using the TBA
\FS.  (As we will explain in section 4,
the running central charge is defined via the Casimir effect on a cylinder.)
It can also be computed in the large-$t$ expansion \CarLud\ and one finds at
first order the same formula as \cR\ with $R=1$. The study of this function
$c(mR)$ for finite $t$, and the search for its corresponding TBA, are the
subjects of this paper and the next.  An immediate result that can be obtained
for finite $t$ is the exactness of $g_{IR}=1-{1\over t}$. This follows most
easily from the $O(n)$ model discretization, and symmetries (see the
discussion of eq.\ 3.20). We therefore see that the massless flow in
sine-Gordon should be observed for $t\geq 1$ only. In the standard
parametrization this corresponds to $4\pi\leq \beta_{SG}^2\leq 8\pi$. In
particular we do not expect this massless flow to have a semi-classical
version since the latter is obtained as $\beta_{SG}\rightarrow 0$.

In the large-$t$ expansion we considered only $\lambda_{\pm 2}$, but
the renormalization group transformations generate higher charges as
well. In the region where $\lambda_{\pm 2}$ are found decreasing, they
disappear for two reasons. On the one hand some pairs of opposite charges
annihilate and contribute simply to renormalizing the coupling constant $g$.
On the other hand pairs of identical charges coalesce to increase first of all
$\lambda_{\pm 4}$. Of course these are irrelevant with respect to the UV fixed
point. Their equation of evolution is made of a damping term $y_4\lambda_{\pm
4}$ and a source term proportional to $\lambda_{\pm 2}^2$. So we expect
$\lambda_{\pm 4}$ to increase driven by the amount of $\pm 2$ charges, and
subsequently to decrease when charges $\pm 2$ have disappeared. The picture
generalizes to all higher charges. Therefore, provided the domains of
evolution of these various charges are separated enough, one can expect more
structure in $c(mR)$ than just a pair of extrema, but it is difficult to make
quantitative predictions.

\newsec{$O(n)$ vertex  model: Symmetries of the thermal flow}

We discuss in this section various properties of the $O(n)$ model on the
honeycomb lattice. This model provides a convenient regularization of our
problem when it flows to the low-temperature ``dense'' phase. It is also
useful in identifying symmetries of the flow and the occurence of level
crossings.

\subsec{$O(n)$ vertex  model and its symmetries}

It is well known that a convenient lattice regularization for the sine-Gordon
model and its restrictions is provided by the $O(n)$ model on the honeycomb
lattice (the links of a hexagonal lattice). The original partition function in
the absence of a magnetic field is
\eqn\zon{{\cal Z}=\sum \beta^M n^L.}
where the summation is taken over all configurations of self-avoiding,
mutually avoiding loops that can be drawn on the lattice, $M$ is the number of
lattice edges occupied by a piece of loop (monomer) and $L$ is the number of
loops. As it stands, \zon\ is not defined locally. A local description is
obtained as follows \Nien. We still associate a weight $\beta$ to each
monomer, but we do not give weights $n$ to loops any more. Instead we put an
orientation on each loop, together with a weight $e^{\pm i\pi e_0/6}$ per left
(right) turn. On a planar honeycomb lattice, the number of left minus the
number of right turns of a closed loop is equal to $\pm 6$. Therefore setting
\eqn\para{n=2\cos\pi e_0,}
we see that summing over all the possible orientations reproduces the desired
weight $n$ per loop, now as a sum of local contributions. We refer to this
model of oriented loops as the $O(n)$ vertex model, or vertex model.  It has
been given other names and is well known to be related in some limits to the
Zamolodchikov-Fateev model and the Izergin-Korepin model;  see
\ref\NienI{B. Nienhuis, Physica 163A (1990) 152;
B. Nienhuis, H. Bl\"ote, J. Phys. A22 (1989) 1415.}.  As we will discuss
below, the correspondence between \zon\ and the vertex model is a little more
intricate in non-trivial topologies like the cylinder or torus.

The difference between the non-local and local formulations is manifest if one
tries to write a transfer matrix. In the first case, ``past'' information
must be taken into account: we need to know if two monomers on a given
column were connected at former times (see \ref\Mar{P. Martin, ``Potts
Models and Related Problems in Statistical Mechanics'', World Scientific,
1991.} and references therein). In the second case, we simply have a vertex
model with an $\check{R}$-matrix acting in the product of two spin-one
representations, and the transfer matrix is the appropriate tensor product of
these $\check{R}$ matrices. To see this consider propagation along one of the
three equivalent directions in the honeycomb lattice. Following
\ref\BaxI{R.J. Baxter, J. Phys.  A19 (1986) 2821.} we decompose the
interactions into elementary three-leg vertices with weights
\eqn\stweight{\eqalign{&\omega_1,\ldots,\omega_7=\beta^{-1},\epsilon,
\epsilon,\epsilon,\epsilon^{-1},\epsilon^{-1},\epsilon^{-1}\cr
&\omega'_1,\ldots,\omega'_7=\beta^{-1},\epsilon^{-1},\epsilon^{-1},
\epsilon^{-1},\epsilon,
\epsilon,\epsilon,\cr}}
where we set $\epsilon=e^{i\pi e_0/6}$.  As usual we can modify these weights
(by a so-called gauge transformation) and leave the partition function
invariant.  The general transformation of interest reads
\eqn\mweight{\eqalign{&\omega_1,\ldots,\omega_7=\beta^{-1},{c'\epsilon\over a},
{a'\epsilon\over b},{b'\epsilon\over c},{a'\over c\epsilon},
{b'\over a\epsilon},{c'\over b\epsilon}\cr
&\omega'_1,\ldots,\omega'_7=\beta^{-1},{a\over c'\epsilon},
{b\over a'\epsilon},{c\over b'\epsilon},
{c\epsilon\over a'},{a\epsilon\over b'},{b\epsilon\over c'}.\cr}}
Let us set
\eqn\qdef{q=-\epsilon^6,}
and  ${a\over b\epsilon^2}=-q$, ${a'\over
b'\epsilon^2}=-q^{-1}$, $c^2=c'^2=ab=a'b'$. We get then
\eqn\fweight{\eqalign{&\omega_1\ldots,\omega_7=
\beta^{-1},\epsilon^{-3},\epsilon^{3},\epsilon^{3},
\epsilon^{-3},\epsilon^{-3},\epsilon^{3}\cr
&\omega'_1\ldots,\omega'_7=
\beta^{-1},\epsilon^{3},\epsilon^{-3},\epsilon^{-3},
\epsilon^{3},\epsilon^{3},\epsilon^{-3}.\cr}}
Let us now stick two such three-leg vertices together. The
rules of the model completely determine the state of the intermediate edge
from the ones of the outer edges, so we can as well forget it to get a
standard four-leg vertex, whose weights determine a $\check{R}$-matrix. This
matrix has two eigenvalues: 0 with degeneracy 6, and $\beta^{-2}-q-q^{-1}$ with
degeneracy three. The respective eigenvectors read
\eqn\eigenI{\eqalign{&|1,1\>\cr
&|1,0\>+q|0,1\>\cr
&|1,-1\>+q|-1,1\>;\ |1,-1\>-q^{-1}|-1,1\>-{a\epsilon\beta\over b'}(1+q^{-2}
)|0,0\>\cr
&|-1,0\>+q|0,-1\>\cr
&|-1,-1\>\cr}}
and
\eqn\eigenII{\eqalign{&|10\>-q^{-1}|0,1\>\cr
&|1,-1\>-q^{-1}|-1,1\>-{a\epsilon\over\beta b'}q^{-1}|0,0\>\cr
&|0,-1\>-q|-1,0\>\cr}}
Clearly this $\check{R}$-matrix has a ``partial'' $U_qsl(2)$ symmetry, where
the generators $S^{\pm}$ only act on the states $|\pm 1\>$ as in a spin
one-half representation (acting on $|0\>$ they just vanish). In the following
we set $S^z=\pm 1,0$. The symmetry $U_qsl(2)$ therefore acts on the loop
degrees of freedom without seeing the empty spaces in the system. By usual
coproduct applications, this symmetry extends to the transfer matrix, up to
boundary effects. Let us emphasize that the $O(n)$ vertex model has more
symmetry than what we are pointing out here. In particular, at the solvable
point of \ref\NienII{B. Nienhuis, Phys. Rev. Lett. 49 (1982) 1062.} there is a
$U_{q'}sl(2)$ symmetry acting on a full spin-one representation, with
$q'=(-\epsilon^6)^{1/4}$. In addition, in the low-temperature limit
$\beta\rightarrow\infty$ there appears a $U_{q''}sl(3)$ symmetry
\ref\Kolya{N.Yu. Reshetikhin, J.Phys. A24 (1991) 2387.} with
$q''=(-\epsilon^6)^{1/2}$. Notice the equality
\eqn\nqrel{n=-q-q^{-1}.}
We restrict in what follows to the case $n\in [-2,2]$ for which $q$ is a
complex number of modulus one.

Let us now discuss boundary conditions. As is well known, the correspondence
between the vertex and the $O(n)$ models breaks down in a non-trivial topology
due to non-contractible loops. For instance, on a cylinder non-contractible
loops in the space direction have a weight 2 in the vertex model because they
have the same number of left and right turns. This can be repaired by changing
the vertex-model boundary conditions. Such an operation is also called putting
a charge at infinity, or a seam \ref\BloN{H. Bl\"ote, J. Cardy, M.P.
Nightingale, Phys. Rev. Lett. 56 (1986) 742.}. Draw on the cylinder a line in
the time direction. Give a vertex configuration an additional weight $e^{\pm
i\pi e}$ every time an oriented edge crosses the line going upwards
(downwards) ($e$ is well defined modulo two). In that case a non-contractible
loop in space direction gets a weight $2\hbox{cos}\pi e$ which can be made
equal to $n$ by the choice $e=e_0$. On a torus there are non-contractible
loops with a more complicated topology. To reproduce their correct weight one
has to consider a family of sectors of the $O(n)$ vertex model characterized
by their ``magnetic charge'' $m=S^z$ (a topological defect $S^z\pi$) and the
``electric charge'' $e$ (a twist term) \nref\Al{F. Alcaraz, M.N. Barber, T.M.
Batchelor, Phys. Rev.  Lett. 58 (1987) 771.} \nref\dfsz{P. DiFrancesco, H.
Saleur, J.B. Zuber, J. Stat.  Phys. 49 (1987) 57.} \refs{\Al,\dfsz}. The
sector with $-e$ and $-S^z$ has identical properties. Let us parametrize
\eqn\para{n=-2\hbox{cos}\pi\gamma,\ q=e^{i\pi\gamma},\ \gamma\in[1,2].}
so in particular $\gamma=(e_0+1)\hbox{ mod }2$.  The various sectors of the
vertex model are then connected by quantum-group commutative diagrams \PS.
This means that for $k$ greater than $S^z=m$, and $k-m$ integer, all levels of
the sector $S^z=k,e=m\gamma$ are contained in the those of the sector
$S^z=m,e=k\gamma$.

\subsec{The critical point}

The $O(n)$ model is critical  when  \Nien
\eqn\betcr{\beta_c=(2+\sqrt{2-n})^{-1/2}.}
Introduce solid-on-solid height variables $\phi$ dual to the arrows, with step
value $\pm\pi$. In the continuum, these degrees of freedom become a free
bosonic field with coupling constant, in the scale where topological defects
are not renormalized,
\eqn\cou{g=\gamma.}
The above parameters $e,m$ are now interpreted as the usual electric and
magnetic charges in a two-dimensional Coulomb gas, with associated conformal
weights $h_{em}(\overline{h}_{em})={1\over 4}({e\over\sqrt{g}}\pm
m\sqrt{g})^2$. The partition function of the vertex model in the $S^z=m,e$
sector, and on a torus of size $R\times T$ reads in the continuum
\eqn\gene{{\cal Z}_{em}\rightarrow
Z_{em}=e^{fR}{1\over \eta(y)\eta(\overline{y})}
\sum_{j\in Z}y^{h_{e+2j,m}}\overline{y}^{\overline{h}_{e+2j,m}};}
for the simple geometry we consider, $y=\hbox{exp}(-\pi
T/R)=\overline{y}$.  The central charge of this vertex model is equal to one
for all values of $n$.

The aforementioned coincidences of levels are easily observed in \gene.  For
example, the ground state of the sector $S^z=m,e=0$ has weight
$h=\overline{h}={gm^2\over 4}$, while weights in the sector $S^z=0,e=m\gamma$
are $h=\overline{h}={(gm+2j)^2\over 4g}$, coinciding with the above for $j=0$.
Such coincidences are well known to occur in the continuum limit as following
from duality in the Gaussian model. We stress that they in fact occur for the
finite lattice model, and for all values of $\beta$.  Notice that $j=0$ is not
necessarily the ground state of this sector: when $m=1$ we have $|2-gm|<g$ so
that $j=-1$ is the ground state for $g>1$.

The partition function of the $O(n)$ model is obtained by proper summation of
\gene\ for various sectors. An important fact is that the ground state of the
$O(n)$ model is not the one of the vertex model. Choosing $e=e_0$ indeed gives
the central charge
\eqn\cen{c=1-{6(g-1)^2\over g}.}
In conformal language, the $O(n)$ model is thus described by a twisted
Gaussian model.  In the $O(n)$ point of view, the value of $c=1$ for the
untwisted Gaussian theory is a $c_{\hbox{eff}}$ value associated with the
presence of a negative-dimension operator.

When $q$ is a root of unity, the $U_qsl(2)$ symmetry implies that additional
levels coincide. We restrict for simplicity to the case
\eqn\beraha{n=2\hbox{ cos}{\pi \over t},\ t=1,2,\ldots.}
and restrict also to sectors $S^z=m$ an integer, where there are an even
number of non-contractible loops in the space direction. In this case the
commutative diagram of \PS\ extends further. Levels in the sector
$S^z=m,e=k\gamma$ (with $k\geq m$) that are not in any other sector are
selected by the restricted partition function
\eqn\geneI{{\cal Z}_{ab}=\sum_{j\in Z}{\cal Z}_{e=k\gamma,m+jt}-{\cal
Z}_{e=m\gamma,k+jt}}
where
$$a=k-S^z,\ b=k+S^z.$$
Differences like \geneI\ are exact for finite lattices, due to the $U_qsl(2)$
symmetry.

Notice that the central charge of the $O(n)$ model reads in the parametrization
\beraha\
\eqn\cmin{c=1-{6\over t(t+1)}.}
The $O(n)$ model at integer $t$ is not the same as the $t^{\hbox{th}}$ minimal
model except for $t=2,3$ ($n=0,1$) \ref\Capitzu{A. Cappelli, C. Itzykson, J.B.
Zuber, Nucl. Phys. B280 (1987) 445.}.  However, in the continuum limit the
above restricted partition functions, as discussed in \PS\ for a similar case
(Potts model), are appropriate pieces of the minimal-model partition functions
\foot{It is necessary to add sectors with $k=1/\gamma$ to restrict the overall
number of non-contractible loops to be even.}
\eqn\contmin{{\cal Z}_{ab}+{\cal
Z}_{a+1/\gamma,b+1/\gamma}\rightarrow
Z_{ab}=\sum_{s=1}^{t}\chi_{as}\overline{\chi}_{bs}}
for $1\leq a,b\leq t-1$, where $\chi$ are the usual characters of the
corresponding irreducible representations of the Virasoro algebra for central
charge \cmin, and the non-universal bulk free energy $f$ has been discarded.
The above expressions sum up to the ones worked out in \dfsz\ to express the
various minimal-model partition functions as linear combinations of Gaussian
ones. For example, the diagonal (i.e.\ $A$-series) partition function reads
\eqn\zmin{Z_{AA}=Z_c\left[{t+1\over t},t\right]-
Z_c\left[{t+1\over t},1\right],}
where $Z_c$ is the Gaussian partition function in the notation of \dfsz. The
expression for finite lattices analogous to \zmin\ also makes sense due to the
$U_qsl(2)$ symmetry. Since this symmetry holds in fact for any $\beta$, all
these expressions also make sense off the critical point.

\subsec{Massless flow}

As discovered in \Nien\ the $O(n)$ model is also critical for $\beta>\beta_c$,
with a continuum limit that is $\beta$-independent. In other words, this
low-temperature phase is a flow to another fixed point, which turns out to
also be a Gaussian model.  This phase is often referred to as the ``dense
phase'' by reference to polymers ($n=0$). Of course the associated massless
flow should be precisely the twisted version of the one discussed in section
2. In particular since the $U_qsl(2)$ symmetry is present all along the flow,
and coincides in the UV and IR with the Gaussian model duality, the relations
between $n$, $g$ and $\gamma$ valid at $\beta=\beta_c$ should still hold true.
This does not require $g$ at the low-temperature fixed-point to be the same as
in the massive phase; as we saw in section 2, we expect the coupling to get
smaller.  Thus another solution of these relations is applicable \Nien; we
have exactly
\eqn\newg{g=2-\gamma=1-{1\over t}.}
The corresponding central charge is given again by \cen\ which reads here
\eqn\newc{c=1-{6\over t(t-1)}}
After proper restriction \refs{\Sm,\Lecl,\FL}, this flow gives rise to the
massless flow from the minimal model at central charge \cmin\ to the one with
\newc.  From the lattice point of view, as emphasized already, all
coincidences of levels in the vertex model transfer matrix hold for every
value of $\beta$. The sectors corresponding to the $O(n)$ model or the minimal
models can be extracted by appropriate combinations of vertex partition
functions. One checks easily that in the dense phase
\eqn\contminI{{\cal Z}_{ab}+{\cal
Z}_{a+1/\gamma,b+1/\gamma}\rightarrow
Z_{ab}=\sum_{r=1}^{t-2}\chi_{ra}\overline{\chi}_{rb},}
so we can see the flow of various sectors of the minimal models:
\eqn\secflow{\sum_{s=1}^{t}\chi_{as}\overline{\chi}_{bs}\quad\hbox{
 flows to }\quad
\sum_{r=1}^{t-2}\chi_{ra}\overline{\chi}_{rb}}
Such a result implies in particular that minimal models flow to a model of the
same kind in the $ADE$ classification \ref\Rava{F. Ravanini, Phys. Lett. 274B
(1992) 345.}.

\subsec{Symmetries in the continuum}

We notice that the above $U_qsl(2)$ symmetry involves the same deformation
parameter as the $U_q\widehat{sl(2)}$ symmetry exhibited in the continuum
theory \sgpert\ since the formal proof for the existence of this symmetry \FL\
does not depend on the sign of $\delta\beta$. We also recall that the soliton
$S$-matrix describing the massive flow exhibits this symmetry as well. To a
large extent one can formally identify the soliton trajectories with the loops
of the underlying lattice model thanks to \nqrel\ \refs{\Zamopol,\Smir}.

In the following we will use results for the continuum as well as the discrete
theory, and usually refer to results in the latter case with an upper index
$``l''$.

\newsec{General numerical analysis: analytic continuation of massive TBA and
lattice data}

\subsec{Analytic continuation}

A natural way of getting insight into the phase structure of the $O(n)$ model
is to determine its ground-state energy $E_0(R)$, where space is a cylinder of
circumference $R$. There are two methods we will use to find this quantity.
The first is to explicitly diagonalize the lattice hamiltonian (or transfer
matrix --- we shall refer to these two interchangeably) on a small,
periodic lattice.  The second way relies on the fact that when spacetime is a
cylinder of radius $R$ and a large length $T$, the partition function is
$Z\sim \exp(-E_0(R) T)$.  If we reverse the roles of space and time, this is
the partition function for a large system at temperature $1/R$.

If the $O(n)$ lattice model were integrable for $\beta^l\ne\beta^l_c$ one
could use the lattice Bethe ansatz to diagonalize the Hamiltonian for a large
number of sites and then do the thermodynamics using the resulting equations
for the energy levels \ref\rYY{C.N. Yang, C.P. Yang, J. Math. Phys. 10 (1969)
1115.}.  Unfortunately, this does not seem possible here. However, this model
is integrable in the continuum. Thus we can use a related but different
method, called the thermodynamic Bethe ansatz (TBA) \ref\rZamTBA{Al. B.
Zamolodchikov, Nucl. Phys. B342 (1990) 695.}. Here one conjectures the exact
$S$-matrix for the particles of the theory, and uses the relation between the
$S$-matrix and the phase shifts of the wave function in order to find the
equations for the energy levels. Like in the ordinary Bethe ansatz case, one
then uses these energy levels to minimize the free energy. The $S$-matrix for
the sine-Gordon model \sgpert\ with $\delta\beta>0$ has been known for a long
time \ZandZ, and the TBA equations were written in \FS. As we discuss in the
sequel, there are complications in understanding possible $S$-matrices and the
TBA for $\delta\beta<0$.  It is thus useful to consider first analytic
continuation of these high-temper\-ature results to the massless phase.  In
this subsection we do this numerically; there seems to be no way of continuing
the TBA equations directly.  In \Zamo\ a similar procedure essentially
reproduced the flow for the ground state energy from the tricritical to the
critical Ising model, a result which was then confirmed by using a ``massless
$S$-matrix''.

Our results can be written in terms of the dimensionless quantity $mR$, where
$m\propto (\delta\beta)^{(t+1)/4}$. (The coefficient of proportionality can be
obtained by comparing the results of conformal perturbation theory and the
TBA.  Similarly, on the lattice $m\propto (\delta\beta^l)^{(t+1)/4}$, where
the coefficient is determined by numerical study.)  We define the quantity
$c(mR)\equiv 6R E_0(R)/\pi$, to take advantage of the fact that in unitary
theories $c(0)$ and $c(\infty)$ are the central charges of the UV and IR
conformal field theories \nref\Affl{I.  Affleck, Phys. Rev. Lett. 56 (1986)
746.}\refs{\BloN,\Affl}. We note that off the critical points this is not the
$c$-function defined in \ZamoI, but at lowest order in $\delta\beta$ the two
are the same.  It is expected (and has been shown in all known unitary cases)
to decrease monotonically as $mR$ is increased.  In non-unitary theories
$c(0)$ and $c(\infty)$ are not the central charges but rather
$c_{\hbox{eff}}\equiv c - 12h -12\bar h$, where $h$ and $\bar h$ are the
lowest conformal dimensions in the sector \refs{\BloN,\Affl}.  Efforts to find
a $c_{\hbox{eff}}$-theorem have used this observation as a starting point.
This makes the ground-state energy a logical candidate, but as we saw in
section 2, it does not always decrease!

The TBA gives a set of integral equations which allow the determination of
$c(mR)$ in the high-temperature phase. They have not been solved directly, but
it is straightforward to solve them numerically. There exists an expansion
around $mR=0$
\eqn\expan{c(z)=c_{UV}+c_{\rm bulk}(z) + \sum_n f_n z^n,}
where we have defined $z=(mR)^{4/(t+1)}$. We also use the variable $r=mR$.
This corresponds to the perturbative expansion of \sgpert\ in powers of
$\delta\beta$ (because of the ${\bf Z}_2$ symmetry $\phi\rightarrow -\phi$,
there are only even powers of $(\delta\beta)^{1/2}$ in the expansion). The
non-universal bulk term does not appear in perturbed conformal field theory
and so is not necessarily an integer power of $z$.  For $t$ integer this bulk
term is the same for the full sine-Gordon theory and for the minimal model,
where it was computed in \Zamo.  This is because the models differ by boundary
conditions only; the truncation does not affect bulk properties. More
precisely one can check numerically (and analytically in related models like
the 8-vertex model) that in the IR the levels of the vertex-model transfer
matrix that belong to the minimal model are still at finite distance in the
$1/R$ scale from its ground state, implying that all these levels share the
same bulk term.

To do the analytic continuation, we first solve the TBA equations for $c$ in
the massive phase to double-precision accuracy. We then subtract off the
massive bulk term. For $t$ even, this term vanishes, but for $t$ odd it is
proportional to $(mR)^2 \ln (mR)$ \Zamo. After subtraction, we have a power
series in $z$ which can be continued to negative values. We do this in one of
two ways, both involving Pad\'e extrapolation.

The first method is to find the ratio of polynomials which go though a set of
$z\ge0$ data points ($z_i, c(z_i)$) determined numerically using the TBA
integral equations. The ratio is unique after one specifies the orders of the
numerator and the denominator; their sum is the number of data points plus
one.  Typically, the most accurate results are obtained by using 10-15 data
points.  We have used the routine in {\it Numerical Recipes}
\ref\numrep{W. Press, B. Flannery, S. Teukolsky and W. Vetterling, ``Numerical
Recipes'' (Cambridge, 1989)}.  It is particularly useful because it provides
a convenient way of estimating the error.  By using different subsets of the
original data points, one obtains different extrapolations. The error is then
estimated by looking at how the result changes for a given value of $z$.

The second method is to find the coefficients $f_n$ in \expan\ by fitting them
to the numerically-determined $c(mR)$ function, using a standard
least-mean-squares program. One then analytically (it is convenient to use
Maple or Mathematica) finds the Pad\'e approximant. (For a polynomial of order
$p$, the Pad\'e approximant is the unique ratio of polynomials of order $p/2$
such that when this is expanded around $z=0$, the first $p+1$ terms reproduce
the original polynomial. If $p$ is odd, then the polynomial in the numerator
is of order $(p+1)/2$ with the denominator of order $(p-1)/2$).  The Pad\'e
approximant should describe $c$ for negative $z$ for much larger values of
$|z|$ than does the polynomial. This method has the disadvantage that it is it
is not as easy to determine the errors involved, and it requires more computer
time, because many more data points are needed for a good fit than for the
first method. For example, in the $t=2$ ($n$=0) case, one finds
\eqn\cexpn{c(z)=1. -.4454536z + .091380z^2 -.00903z^3 - .00018z^4 + .00015z^5,}
where the numerical error is in the last digit given.  The resulting Pad\'e
approximant is
\eqn\cpade{c(z)={1 - .301756 z + .0436 z^2- .00314 z^3\over
1 + .143698z + .0162z^2}.}
We have checked for several values of $t$ that the results obtained by these
two methods are consistent, but we have mainly used the first method.

It is easy to find the first extremum of the $c$-function using this
continuation. For large $t$ this allows us to check formula \cext\ for this
maximum, which is determined by quite a different route. Before making the
comparison, we must subtract off the low-temperature bulk term. The situation
here is a little more complicated than for the massive flow.  For $0< n\leq 2$
the bulk term is the same for the full sine-Gordon model and the minimal
theory, and we can use the results of \Zamo. This is not true for $n=0$
($t=2$) due to a large number of level crossings which we shall discuss
further.  For the sine-Gordon bulk term, our results are still consistent with
putting $t=2$ in the minimal-model formulas.  The agreement is very good, as
exemplified in the following table that gives the ratio of $c_{max}-1$ for the
continuation and \cext
\medskip
\settabs 8 \columns
\+&t=2&t=3&t=4&t=5&t=6&t=7&t=8\cr
\+&.556&.685&.759&.806&.840&.87&.88\cr
\medskip
\noindent
We certainly see the $1/t^3$ behavior of the maximum being reproduced in the
continuation, and standard extrapolation as a function of $1/t$ shows very good
convergence to the value one.  As $t$ increases the number of coupled
equations in the massive TBA increases, making the calculation more difficult,
as well as the effect smaller to see. Other features of the large-$t$
expansion can also be checked against the TBA results, like the ratio of the
$r$ values at the maximum and at the intersection with $c=1$.

We have not been able to extend the continuation far enough to observe a
minimum, nor $c$ going back to one in the IR. We give an example of data
obtained in figure 1 for $n=\sqrt{2}$.

It is difficult to determine precisely the radius of convergence of the $z$
expansion for $c(mR)$, but several tests provide a value around $z_c\approx
1$. The Pad\'e extrapolation typically gives reliable data up to $z\approx 3$
for $n=0$, and less as $n$ increases.  Singularities in the complex plane can
be localized using the Pad\'e approximants, but their position depends too much
on the number of points used to be reliable. There was however no sign of
these singularities getting too close to the real axis, negative or positive.
(If there were a pole on the negative axis, the terms in \cexpn\ should
alternate in sign.)

\subsec{Lattice results}

A complementary tool for studying the massless flow is the numerical
diagonalization of the $O(n)$ vertex model transfer matrix on cylinders of
lattice width $R^l$. (Since we will use results for the continuum as well as
the discrete theory, we often refer to results in the latter case with an
upper index ``$l$''.)  This provides lattice data that, after proper
extrapolation $|\beta^l-\beta_c^l| \rightarrow 0$, $R^l\rightarrow\infty$ with
$r^l\equiv|\beta^l-\beta_c^l|^\nu R^l\ $ fixed, should reproduce the continuum
limit of interest ($r$ and $r^l$ being related by a non-universal
proportionality constant). A difficulty here is to extract the bulk term.
Several obvious strategies are possible, none of them giving too-satisfactory
results. Their comparison however allows the determination of $c(r^l)$ with a
reasonable accuracy, generally better for small or large values of $r^l$ and
for $t$ large.  A convenient number to consider is
\eqn\latc{c^{(R^l)}(r^l)={{\log\Lambda^{(R^l_1)}(m^l)\over R^l_1}-
{\log\Lambda^{(R^l_2)}(m^l)\over R^l_2}\over {1\over (R^l_1)^2}-
{1\over (R^l_2)^2}}.}
where $R^l_1$ and $R^l_2$ are neighboring widths (eg $R^l,R^l-1$). In the
study of critical points such combination is known to converge very rapidly to
the exact value $c$, with the bulk term conveniently extracted. In our
situation, we are not allowed a priori to use this procedure, since apart from
a regular term, the bulk term (actually its singular part from the lattice
point of view, but that is all what is seen in the continuum limit) also
scales like $1/R^l$ in the limit we are interested in. On the other hand,
corrections to scaling for finite $R^l$ seem to compensate conveniently so
that \latc\ is a rather reliable estimate, especially at $t$ large, i.e.\ it
coincides with estimates obtained by other, more correct means, up to error
bars. As an example we give also in figure 1 the result obtained from \latc\
for $n=\sqrt{2}$.  After proper rescaling of the lattice and continuum
variables, it is seen to nicely coincide with the result from TBA
extrapolation in the region where they are reliable. For another example we
give in the following table the maximum of the function of $c^{(R^l)}$ (for
$R^l_1=R^l,\ R^l_2=R^l-1$) for various sizes
\medskip
\settabs 7 \columns
\+&$R^l$=&4&5&6&7&8\cr
\+&$c_{max}$=&1.024&1.017&1.016&1.015&1.015\cr
\medskip
\noindent
which seem to be converging nicely to the TBA value $c_{max}=1.0137$. In the
lattice simulation (but {\bf not} the TBA) one observes further $c$ reaching a
minima, and going back to the value 1 in the IR. Details of such curves can be
checked against the large $t$ results, like the value of $c$ at the minima,
the ratio between the values of $r^l$ at the maxima and minima, or at the
maxima and the point $c=1$, with very reasonable agreement.

Two additional features must be mentioned. First we have observed a general
tendency for the TBA and lattice data to differ after some value of $r^l$, at
the border of the domain where the TBA data are reliable. This is observed in
particular for the unsubtracted case, where lattice data are affected only by
the small uncertainties of the $R^l\rightarrow\infty$ extrapolation.  In
addition, we have observed a large unstability of the lattice data in the
region $r^l$ large.  This is very marked for $n<0$ in particular, where for
$R^l$ large enough we have observed another pair of extrema in \latc\ as well
as related features in the unsubtracted data. An example is given for $n=-1$
in figure 2.

\newsec{The boundary case $n=-2$}

The case $n=-2$ is easily solvable.  Indeed, the associated Gaussian model at
the UV fixed point has $g=2$, and in the Thirring version, the four-fermion
coupling vanishes. This is actually the ``natural'' boundary of the behavior
we want to study since the dense $O(n=-2)$ model, corresponding formally to
$g=0$ in \newg , is in fact massive.  So we shall not be able to recover a
Gaussian model in the IR, and this example is pathological in some
respects.

That we are dealing with free fermions can also be seen directly from the
point of view of the $O(n=-2)$ lattice model.  Consider indeed the fermionic
integral
\eqn\zfermi{{\cal Z}=\int d\psi d\psi^+ \prod_i e^{-\psi_i\psi_i^+}
\prod_{\<jk\>}
\hbox{exp}\left[\beta^l(\psi_j^+\psi_k+\psi_k^+\psi_j)\right].}
By usual rules of Grassman integration this gives rise to an expansion similar
to \zon\ with the only difference that ``loops'' covering a single edge
(corresponding to terms like $\psi_j^+\psi_k\psi_k^+\psi_j$) are allowed here.
Such a model of loops + dimers \ref\Mat{I. Kostov, M. Staudacher, Nucl. Phys.
B384 (1992) 459.} is expected to be in the same universality class as \zon.
On the other hand it is exactly solvable by elementary techniques since
\zfermi\ is equal to the determinant of the interaction matrix. The critical
point occurs at $\beta^l_c=1/3$ where this matrix restricts to the discrete
Laplacian.  For $\beta$ close to $\beta_c$ (mapping between lattice and
continuum is straightforward here, so we suppress the lattice label) the
matrix coincides with $\Delta+m^2$ with $m^2=9(\beta_c-\beta)$.  Therefore the
massive flow corresponds to a positive $m^2$ while the massless flow
corresponds to negative $m^2$. These results are in agreement with the value
$h=\bar h=0$ for the thermal operator of the $O(n=-2)$ model. The sector where
non-contractible loops in the spatial direction have weight 2 is reproduced by
choosing antiperiodic boundary conditions for the fermions on the cylinder.
In the massive flow the running central charge is immediately obtained from
\nref\IS{C. Itzykson, H. Saleur, J. Stat. Phys. 48 (1987) 449.}
\nref\KM{T. Klassen, E. Melzer, Nucl. Phys. B350 (1990) 635.} \refs{\IS,\KM}:
\eqn\ceff{\eqalign{c=1-&{3r^2\over \pi^2}\left(\hbox{ln}{\pi\over r}
+{1\over2}+-\gamma_E\right)\cr +&{12\over\pi}\sum_{k=1}^\infty\left[
\sqrt{(2k-1)^2\pi^2+r^2}-(2k-1)\pi-{r^2\over 2(2k-1)\pi}\right],\cr}}
where $r=mR$. In the standard $O(n=-2)$ model this result is still expected
to hold by universality, although the correspondence between the mass $m$ and
the lattice temperature may involve additional logarithmic contributions.

Consider now the massless direction. The above formula is still expected to
hold, up to a bulk $r^2$ (and possible logarithmic corrections) term.  The
striking fact is then the presence of the square root singularities which,
beyond the value $r^2=-(2k-1)^2\pi^2$, develop an imaginary part. (Similar
square-root singularites seem to appear in affine Toda theories at negative
$z$ \KM). Recalling the expression for the continuum partition function
(including massive states as well as the ground state) of the $O(n=-2)$ model
in that particular sector, where $T$ is the length of the cylinder,
\eqn\zn{\hbox{exp}\left(4\pi {T\over R}{c\over
24}\right)\left\{\prod_k \left[1+\hbox{exp}\left(-{T\over
R}\sqrt{(2k-1)^2\pi^2+r^2}\right)\right]^2+(-)^2\right\},}
we see for instance beyond $r^2=-\pi^2$ the two largest eigenvalues become
equal in modulus. Successive branchings occur as $r^2$ crosses the roots
$-(2k-1)^2\pi^2$.  It is not clear how to define a central charge when
the largest eigenvalue becomes complex. The most natural candidate is the real
part of \ceff\ (the modulus of the largest eigenvalue). Because of the
square-root term, it is singular with an infinite derivative at each of the
points $r^2=-(2k-1)^2\pi^2$, and shows wide oscillations moving away from the
UV fixed point. This is fully confirmed by the numerical study of the lattice
$O(n=-2)$ model \zon.

It is interesting to compare this situation with the square-lattice Ising
model. Introducing the variable $v=\hbox{tanh}\beta/2$ the locus of zeroes of
the partition function is a set of two circles depicted in figure 3.  The
ferromagnetic critical point is the intersection of the rightmost circle with
the positive real axis. Away from the critical point the Ising partition
function is a sum of square roots of determinants of the massive Laplacian.
The case $m^2$ real and positive corresponds to moving on the real physical
axis (in which direction depends on the sign of $m$). The case $m^2$ real and
negative corresponds to moving exactly along the circle of singularities. The
intersection of the two circles corresponds to $m^2=-4$ which is also the
opposite of the maximum eigenvalue of the massless Laplacian on the square
lattice.

That the largest eigenvalue of the transfer matrix in a sector can branch
and become complex for real $\beta$ seems peculiar to $n=-2$ \ref\BN{H.
Bl\"ote, B. Nienhuis, Physica A160 (1989) 121.}. For any other $n\in [-2,2]$
and for $R^l$ large enough this eigenvalue appears to remain real in our
computations.  It must be so in the $r^l\rightarrow\infty$ limit so that it
agrees with the known spectrum in the continuum.

Although pathological, the case $n=-2$ is interesting since then the running
central charge presents an infinite sequence of singularities and
oscillations. This is a strong indication that for $n$ negative and close
enough to $n=-2$ similar oscillations are present (a fact we observed
numerically above for $n=-1$) and maybe also some singularities. The region
$n$ negative seems more difficult to study because of these features, and in
the remainder of this paper and the sequel we mainly consider $0\leq n\leq 2$.

As a final comment, we note that the general case should also be amenable by
exact solution (using the Bethe ansatz) of the Thirring model with imaginary
mass. We have not done this calculation yet.

\newsec{Detailed study of the case $n=0$: \nl
Spontaneous breaking of $N$=2 supersymmetry}

We now discuss in some detail the case $n=0$. It is closely related to the
physics of self-avoiding walks and also to $N$=2 supersymmetry
\ref\S{H. Saleur, Nucl. Phys. B382 (1992) 486.}. Again, we analyze the
ground-state energy.  It was shown in \ref\sgs{G. Waterson, Phys.  Lett. B171
(1986) 77} that the sine-Gordon model \sgpert\ is $N$=2 supersymmetric at the
coupling where $n$=0. This will prove a very powerful tool. The fermion number
$F$ in the $N$=2 language is equal to half the soliton number in the
massive sine-Gordon model.

The results depend crucially on the boundary conditions around the cylinder
\FS. These are implemented by giving non-contractible loops a weight $2\cos\pi
e$ in the $O(n)$ model, which corresponds to a fugacity for
the sine-Gordon solitons of $\exp{\pm ie \pi}$. In the $N$=2 picture, this
corresponds to inserting the operator $e^{i2\pi e F}$. The boundary conditions
used in sect.\ 4 have $e$=0, and in the $N$=2 language these are called
Neveu-Schwarz.  These give an effective central charge $c=1$ in the UV limit.
We first discuss the situation with twisted boundary conditions, which have a
``charge at infinity'' $e=e_0=\pm 1/2$. This gives non-contractible loops a
weight of zero, and has $c_{UV}=0$.  In the $N$=2 picture this means that we
insert $(-)^F$, and is called the Ramond sector. These boundary conditions
have the advantage that we can derive quantitative information by using the
supersymmetry, even in the massless phase.

\subsec{The Ramond sector}

\noindent
{\rm a) Preliminary: one-dimensional phase transitions for polymers}
\smallskip

There are several aspects that deserve attention. As a preliminary we recall
that for lattice cylinders of finite circumference $R^l$, the flow to the
dense phase is characterized by an infinite sequence of ``first-order'' phase
transitions, taking place at values of $\beta^l>\beta^l_c$ which depend on
$R^l$.  These transitions are easily understood by studying the correlation
functions of $L$-leg operators, which are defined by ``forcing'' a set of $L$
lines to propagate on a cylinder (see figure 4) with as usual a weight
$\beta^l$ per edge. These operators have UV conformal dimensions $h=\bar
h=(9L^2-4)/96$. The resulting partition function grows like the largest
eigenvalue $\Lambda^{(R^l)}_L$ of the corresponding transfer matrix. For
$\beta^l\leq\beta^l_c$ self-avoiding walks cover a negligible fraction of the
lattice, so this eigenvalue must be smaller than one. If the total number of
possible configurations for this set of $L$ lines of total number of edges
${\cal N}$ grows as $\Omega^{(R^l)}_L\propto\left(\mu^{(R^l)}_L\right)^{{\cal
N}}$, they will start occupying a finite fraction of available space when
$\beta^l={1\over \mu^{(R^l)}_L}$. In other words, this eigenvalue of the
transfer matrix will cross the previously-dominant trivial eigenvalue, which
is always equal to one for any $\beta^l$. The polymer density is discontinuous
at that point, and this crossing can be described as a first-order
one-dimensional phase transition \ref\BT{R. Balian, G.Toulouse, Ann. Phys. 83
(1974) 28.}.

To illustrate this point, it is clear that the transfer matrix for propagation
of two (unoriented) lines on a cylinder of radius $R^l=2$ is one-by-one, with
a single eigenvalue equal to $2(\beta^l)^4$. The latter crosses one for
$\beta^l=2^{-1/4}$ \ref\thesis{H. Saleur, Ph.D. Thesis, Paris 6 University.}.

For fixed $L$, $1/\mu^{(R^l)}_L\rightarrow\beta_c^l$ as
$R^l\rightarrow\infty$. Moreover, by the usual finite-size scaling arguments,
$(1/\mu^{(R^l)}_L-\beta_c^l)^{\nu}\propto 1/R^l$.  Therefore in the
``blown-up'' variables of the continuum limit, the transitions occur at finite
distance from one another. We introduce the notation
\eqn\rL{\lim
_{R^l\rightarrow\infty}(1/\mu^{(R^l)}_L-\beta_c^l)^{\nu}R^l=r^l_L.}
A set of estimates of $r^l_L$ obtained from lattice computations is
given in the following table.

\medskip
\settabs 7 \columns
\+&$R^l$=L&$R^l$=L+1&$R^l$=L+2&$R^l$=L+3&$R^l$=L+4\cr
\smallskip
\+$L=\,$2&.8101&.779&.759&.746&.738\quad $\longrightarrow$&\quad .725\cr
\smallskip
\+\quad\quad 4&1.920&1.788&1.725&1.681&1.653$\quad
\longrightarrow$&\quad 1.6\cr
\smallskip
\+\quad\quad 6&3.032&2.811&2.700&2.63&2.55\quad $\longrightarrow$&\quad 2.2\cr
\smallskip
\+\quad\quad 8&4.147&3.85&&&&\cr
\medskip

\noindent
where the last column is the extrapolation to $R^l\rightarrow\infty$. They are
universal up to an overall scale.

\bigskip
\ifx\answ\bigans\else{\vfill\eject}\fi
\noindent b) Level crossing in the Ramond sector
\smallskip
By using the quantum-group symmetries discussed in sect.\ 3, one easily finds
that the eigenvalues describing $L$-leg operators are also present in the
Ramond sector ($e=1/2$) for $L=2\hbox{ mod }4$. One can see this explicitly
for the value $R^l=2$, where the transfer matrix of the $O(n=0)$ vertex model
in the sector of vanishing spin reads
\eqn\tmat{\left(\matrix{(\beta^l)^4&-i(\beta^l)^4&0\cr
                        i(\beta^l)^4&(\beta^l)^4&0\cr
                        2(\beta^l)^2i^{1/2}&2(\beta^l)^2i^{-1/2}&1\cr}
\right).}
The eigenvalues are $1,2(\beta^l)^4,0$. Hence  at $r^l_{2+4n}$ a new
eigenvalue crosses the value 1, which of course is the continuation of the
largest eigenvalue in the high-temperature phase. This crossing is a
first-order phase transition for finite $R^l$ and remains so in the
$R^l\rightarrow\infty$ limit, as can easily be checked using scaling or by
numerical analysis.

Much can be learned about these first-order phase transitions from the $N$=2
point of view. Well-known arguments \ref\Witten{E.  Witten, Nucl. Phys.  B202
(1982) 253.}\ show that the Ramond ground-state energy being zero is
equivalent to supersymmetry remaining unbroken (for unitary theories).  Since
the Witten index is not zero here, supersymmetry cannot be broken in the
high-temperature phase where the model is unitary, and indeed the largest
eigenvalue of the transfer matrix is one. However, unitarity is lost in the
low-temperature phase, and it is possible for a level to cross zero energy
(the continuation of the Ramond ground state) and become negative.  This is
actually easy to see using the $N$=2 Landau-Ginzburg action \ref\rLG{A.B.
Zamolodchikov, Sov. J.  Nucl.  Phys. 44 (1986) 529;\nl D.  Kastor, E.
Martinec, S.  Shenker, Nucl.  Phys.  B316 (1989) 590; \nl E.  Martinec, Phys.
Lett. 217B (1989) 431;\nl C.  Vafa, N.P.  Warner, Phys. Lett.  218B (1989)
51.} with superpotential
\eqn\superp{X^3+\overline{X}^3+(\delta\beta)^{1/2}(X+\overline{X}),}
For $\beta>\beta_c$ the potential is not positive (or even real), ensuring
that the ground-state energy does not have to be zero.  We can say that
supersymmetry is spontaneously broken in the Ramond sector for $r^l>r^l_2$,
and that at each $r^l_{2+4n}$ a new level crosses energy zero. Notice however
that, since the perturbation does not break supersymmetry explicitly, the
Witten index remains equal to its UV value, and in fact a pair of levels with
opposite values of $(-)^F$ cross zero together.

If one simply measures the ground state energy of the $R$ sector, it is zero
up to $r^l_2$ and then non-zero with a discontinuity of the first derivative
at that point. However there is also a discontinuity of the bulk term, so
if one measures running ground-state energies after subtracting this
bulk term, the final result has a discontinuity at $r^l_2$, as is illustrated
in figure 5 from lattice data.  This is an indication that the analytic
continuation done in the last section only describes the ground state energy
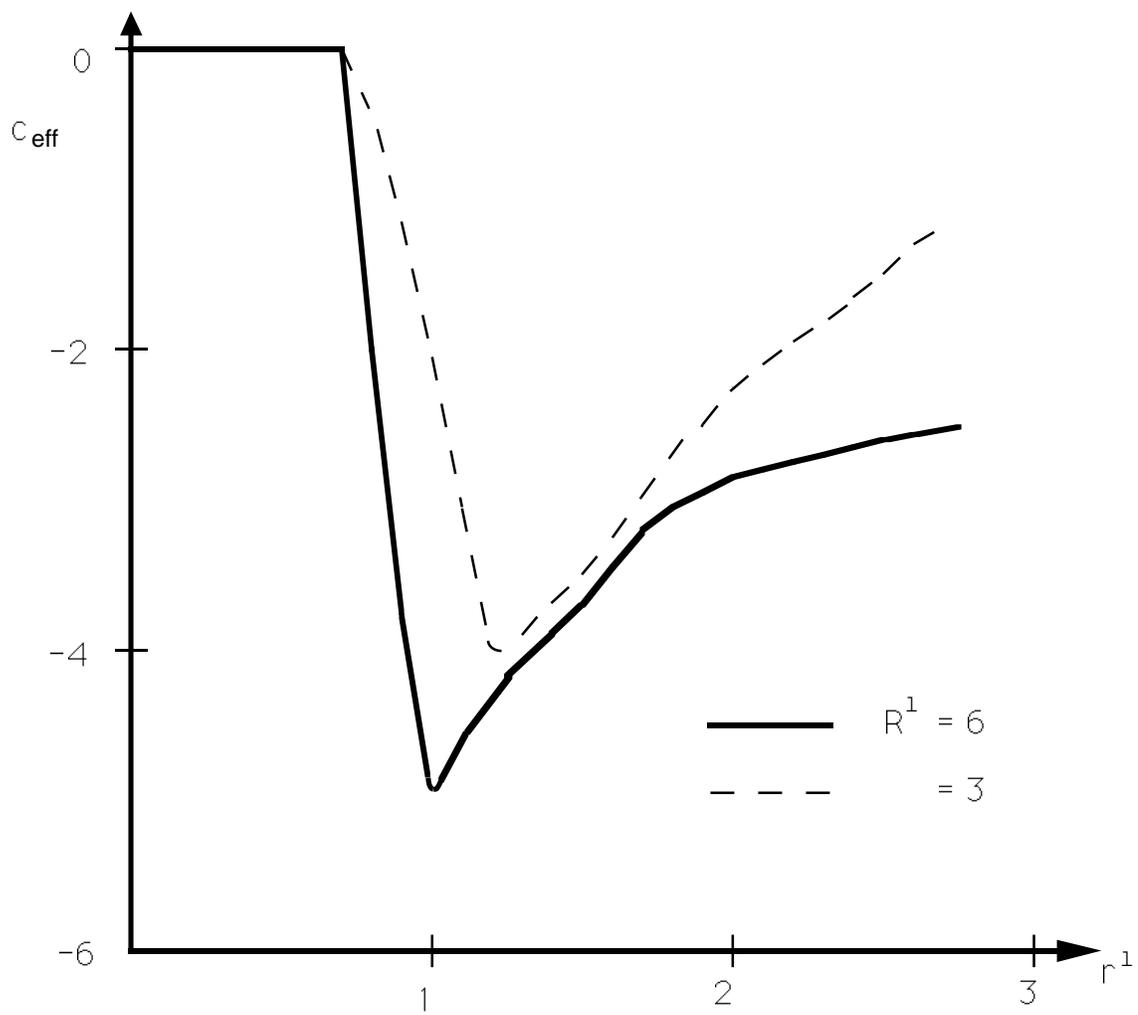
up to $r_2^l$: since $c=0$ in the Ramond sector for $z\geq 0$, analytic
continuation obviously gives $c=0$ for all $z$. Of course all the above
results apply without reference to the lattice model, and using the variable
$r$.

This picture can be further elucidated by considering the ``index'' studied in
\refs{\CV,\CVFI}, $Q=i(R/T)\, \tr\,e^{-R H}F(-)^F$. \foot{Notice that this
Hamiltonian propagates the system in the $R$-direction, as opposed the
transfer matrix we studied above, which is in the $T$-direction.} The index
$Q$ is obtained by taking the derivative of the ground state energy with
respect to $e$ at $e=1/2$. It has a simple polymer interpretation in the
massive flow: it is the partition function for a single non-contractible loop
\FS.  This index has the advantage that we can analytically derive an
differential equation for it even in the massless phase, as opposed to the
ground-state energy in the Neveu-Schwarz sector discussed in the last section,
where the TBA integral equations can only be numerically continued. (If one
tries to continue the equations directly, the integrals diverge.) Since all of
the derivations of \CV\ hold for any superpotential, they should hold even in
the case of a non-real potential like \superp\ in the dense phase. It was
shown in \refs{\CV,\CVFI}\ that in the massive phase $Q$ satisfies a
differential equation of Painlev\'e III type: we have
\eqn\PIII{{d^2U\over dr^2}+{1\over r}{dU\over dr}=\hbox{sinh }U,}
with
\eqn\Q{Q={r\over 2}{dU\over dr}.}
Near $r\equiv mR=0$ the solution we are interested in looks like
\eqn\asym{U\sim {2\over 3}\hbox{log }{r\over 2}+ s+ \sum_n g_n r^{(4n/3)},}
where $\hbox{exp}(s/2)=\Gamma(1/3)/(2^{2/3}\Gamma(2/3))$ \ref\MTW{B.M.  McCoy,
C.A. Tracy, T.T Wu, J. Math. Phys. 18 (1977) 1058.}.  As before, we define
the variable $z=r^{4/3}$. The low temperature phase $\beta >\beta_c$
corresponds to $z$ negative but real. Defining $u=U+i\pi/2$, \PIII\ is then
\eqn\PIIIz{{d^2u\over dz^2}+{1\over z}{du\over dz}=
{9\over 16\sqrt{-z}}\hbox{cosh }u.}
It follows from \asym\ that our solution for $u$ and $Q$ is real for negative
$z$.

For $r$ near $r^l_{2+4n}$ two eigenvalues of the transfer matrix (two levels
of the hamiltonian) are very close for $e=1/2$, and are coupled if $e\sim 1/2$
by the standard formula of second-order perturbation theory. It is easy to
show that the numerator is of order $e-1/2$ while of course the denominator is
the difference of energies of the two levels. Therefore we expect the
derivative with respect to $e$ to exhibit a simple pole at $r_{2+4n}$, that is
we expect {\bf simple poles} for $Q$. That $Q$ should become infinite when a
level crosses zero energy is clear a priori from its definition where no bulk
term contribution is subtracted. In terms of polymers this pole corresponds to
the sudden growth of infinite arms to the non-contractible loop (figure 4b).
This can be seen explicitly on the lattice for radius $R^l=2$, where for $e$
close to $1/2$ the eigenvalue that is dominant in the UV reads
\eqn\noname{1+\left(2i\beta^4-{4i\beta^6\over 1-2\beta^4}\right)\pi(e-1/2).}

Hence $Q$ should exhibit an infinite sequence of simple poles in this massless
flow. A numerical solution of \PIIIz\ fully confirms this prediction. The
locations of the first poles (in the variable $r$) are
\medskip
\settabs 6\columns
\+&$1^{st}$&$2^{nd}$&$3^{rd}$&$4^{th}$&$5^{th}$\cr
\smallskip
\+&2.95708396&8.61&14.1&19.5&24.8\cr
\medskip

\noindent
Taking ratios to eliminate the lattice-dependent mass scale they agree well
with the above results of lattice computations. For the reader with this
particular interest we give independently in the appendix the results for
polymers deduced from this study.

In fact, we can extract information about these poles analytically. The first
pole $z_1$ can be determined from the coefficients $g_n$ in
\asym: as long as there are no other poles nearby, the coefficients should be
the same as those of the expansion of $1/(z-z_1)$ in powers of $z$. The $g_n$
can be found analytically by substituting \asym\ into \PIII; for example,
$g_1= -(9/32)2^{2/3} e^{-s}$. Alternatively, one can fit these coefficients to
a ratio of polynomials as in sect.\ 4, and find the first pole in this manner
(this is why the location of this pole is known very accurately).

It is easy to show that all poles in $Q$ are simple ones. Near a pole, the
cosh-Gordon equation \PIIIz\ reduces to the (radial part) of the Liouville
equation (i.e.\ $\cosh u\sim e^u/2$). This can be solved explicitly; the
general solution near a pole at $z=z_0$ is
\eqn\Liou{u(z)\sim \ln {-64\over 9 \sqrt{-z}}{z_0\over (z-z_0)^2}}
Thus all poles in $Q \propto du/dz$ are simple.

We can also see that there are an infinite number of poles on the negative
$z$-axis. We show that given one pole, there is another one a finite distance
farther along the axis.  The crucial fact is that in any given region, the
right-hand side of \PIIIz\ is finite and positive. (It may appear more
comfortable to suppress the $1/\sqrt{-z}$ by switching to the variable
$w=(-z)^{4/3}$, where this factor disappears and the right-hand side is $\ge
1$.) As we move away from the pole in the negative direction, $u''$ and $u'$
are positive.  As long as $u'/z$ is negative, \PIIIz\ requires that $u''$ is
positive and finite. Thus $u'$ must decrease at a finite rate and eventually
hit zero.  This is a minimum of $u$, because on the negative $z$-axis, $u$ has
no maxima (only poles): if $u'=0$, $u''>0$.  Thus after the minimum, $u$ must
start increasing and not turn around. It must increase at a finite rate, so
eventually $u$ will get large enough so that we can use replace \PIIIz\ with
the Liouville equation, and our solution is of the form \Liou.  Since $u'$ is
negative and must remain negative until we hit a pole, the pole in $\Liou$
must fall on the negative axis. Thus we have established that there is another
pole a finite distance away at larger $|z|$.

In fact, for $|z|$ large we can show that the distance between a pole at
$z_i$ and the next pole is $\sim C|z_i|^{1/4}$, where $C$ is a constant which
is about $7$. This is done by patching together the solution
\Liou\ for $u$ large with the solution for $u$ small (i.e.\ $\cosh u\sim 1$),
which is
$$c(z)={1\over 4}z^{3/2} + A \ln z +B.$$

As $r\rightarrow\infty$ an infinite number of levels crosses the value one.
The Witten index $\tr(-1)^F$ is still equal to 2. However, most levels other
than the ones of vanishing energy have a bulk term that becomes infinite in
that limit. If we subtract this bulk term, we have to set the Witten index to
zero for consistency. This is in agreement with the study of the IR fixed
point described by an $\eta\xi$ system. We have here two theories related by a
$N$=2 preserving flow which have different Witten indices, so they must be
considered ``at infinite distance'' from each other \ref\CK{M.  Cvetic, D.
Kutasov, Phys. Lett. B240 (1990) 1502.}. This sheds some doubt about the
possibility of connecting them smoothly by a TBA-like description, an issue
which we will discuss further in the next paper.

Besides the crossing of eigenvalues at $r^l_{2+4n}$ just explained by
quantum-group symmetries, we have observed additional crossings at values that
tend toward $r^l_{4+4n}$ for large lattice systems. The matrix elements are
such that these crossings do not produce a pole in $Q$.

\subsec{The Neveu-Schwarz sector}

\noindent a) Observations
\smallskip
This is the sector discussed in sect.\ 4, where solitons for the massive flow
in the sine-Gordon model have fugacity one, and non-contractible loops in the
$O(n)$ vertex model are given a weight 2. Unfortunately, there is no index in
this sector and the TBA continuation of sect.\ 4 breaks down after the first
maximum. We here discuss only the lattice results.  As in the large-$t$
expansion, $c=1$ is observed both in the UV and IR. A remarkable feature, very
different from the massive study, is the very slow stabilization of the
curve's shape as the lattice size is increased, and one may wonder if this
hides further aspects, unobservable at the sizes we can study. However, a
clear result which generalizes to all values of $n>-2$ is that the largest
eigenvalue of the transfer matrix remains real and non-degenerate at all
values of $r^l$.  For the moment we start by doing some ``numerology''.
Comparison of estimates for $r_L^l$ with figure 6 shows that peaks of the
derivative ${d^2c\over d(r^l)^2}$ occur roughly at $r^l_2$ and $r^l_4$. These
coincidences are confirmed by a more detailed study, and they hold within our
error bars. At $r^l=r^l_2$ there occured a phase transition in the Ramond
sector, associated with a discontinuity of the running effective central
charge. It is not surprising that $r^l_2$ emerges as well in the study of the
$NS$ sector. Even though we are a finite distance from the UV fixed point in
the blown-up variable $r^l$, we are still infinitesimally close to it,
and the polymers present in the ground state of this sector still occupy
an infinitesimally small fraction of the available space. This does not change
the fact that at $r^l_2$ two-leg diagrams become infinite.  One may actually
wonder whether $r^l_2$ (as well as $r^l_{2n}$) could be a point of phase
transition in the sine-Gordon model itself, independent of the sector; we
shall discuss this further.

We have also observed that the two (first) extrema of the curve $c(r^l)$ occur
roughly at the values $r^l_1$ and $r^l_3$. Study of the present data indicates
that they do not exactly take place there, although as always this may be a
finite-size effect. We do not expect any special feature to occur in the
Ramond or Neveu-Schwarz sector at values $r^l_{1+2n}$. This is because the
transfer matrix we are considering acts on spin zero in the vertex model
language, so only even numbers of lines can propagate through the system.
\bigskip
\noindent b) Possible singularities
\smallskip
The evidence seems to exclude a discontinuity in $c(mR)$. As far as its
derivatives are concerned, they are difficult to study numerically. The
lattice-subtracted quantity \latc\ is not very reliable for determining
derivatives, nor are any of the procedures we have used to subtract the bulk
term. We can of course study derivatives of unsubtracted quantities, but then
their convergence as $L$ increases is affected by $c$ and the bulk term, and
this mixture is difficult to disentangle. The safest quantity to study seems
to be the second derivative with respect to $r$ of the full ground-state
energy, including the bulk contribution. In the large $R^l$ limit, the bulk
term should simply contribute an additive constant. Results are shown in
figure 7.  There is a spectacular peak at values that converge to $r^l_2$. The
convergence past this value is however quite slow, and makes the quantitative
analysis of this peak difficult. We have done the same analysis for the Ramond
sector, where we know that we should observe a delta function. The numerical
results look virtually identical to these past $r_2^l$. We therefore suggest
that there may be a {\bf discontinuity of the first derivative of $c$} in the
Neveu-Schwatz sector at $r_2$. The analysis of $r_4$ shows similar but less
marked properties. We could not get any information about other $r_{2n}^l$.
However if there is indeed a singularity for $r_2$ and $r_4$ it seems natural
there is one at all $r_{2n}$.  Recall that in the polymer language, at
$r=r_{2n}$ it becomes possible for $2n$ lines to propagate along the cylinder
in time direction, and that this interferes with the characteristic feature of
the ground state in the UV limit, the domination by non-contractible loops in
space direction (with weight 2).

To investigate further the possibility of a singularity at $r^l_2$ we studied
in detail the structure of the transfer matrix in the Neveu-Schwarz sector for
complex $\beta^l$. For finite lattice size $R^l$, the largest eigenvalue does
not cross for $\beta^l$ real. One can however locate square root branch points
in the complex plane, a pair of which moves (in the blown-up variable $r^l$)
towards $r^l_2$.  The situation here is somewhat simpler than for ordinary
critical phenomena, since in the variable $r^l$ the infinite number of branch
points that converge to the critical inverse temperature $\beta^l_c$ are now
at finite distance from each other. Hence we have to consider the limit point
of a single pair of branch points, which is expected to exhibit a singularity
of the form $x^k|x|$ (see \ref\Wood{D.W. Wood, J.Phys. A 20 (1987) 3471; D.W.
Wood et al., J.Phys. A 20 (1987) 3495.} for more detailed discussion). This
agrees with the above expectation of $k=0$. However a singularity of the
largest eigenvalue is not enough here; we need to make sure that it occurs in
its scaling part. We unfortunately did not get definitive numerical evidence
for this.

\subsec{Miscellaneous}

We have two additional observations. At the UV and IR fixed points, we have
$h=gm^2/4$ in the NS sector. Thus we can define the running coupling constant
$g(r^l)$ by measuring the gap between the ground state of the NS sector and
the ground state of the sector where two non-contractible loops propagate
($m$=2). The result for available sizes is a monotonic curve with a derivative
maximum at a value close to $g=1$, close also to $r^l_4$ (these coincidences
do not seem to be exact).

We have also investigated whether the model passes ``close'' to an
intermediate fixed point by studying the spectrum of excited states and
looking for the tower structure characteristic of a conformal field theory. No
such structure has been observed.

\newsec{Conclusions}

We have explained why $n$ negative was expected to exhibit even more
complicated properties, and we shall not discuss it further. For $n\geq 0$,
the key features of the large-$t$ expansion are reproduced by the various
numerical studies. In particular, we emphasize that despite the complex nature
of the interaction, the ground state energy of the sine-Gordon model (except
for $n=-2$) remains real all along the flow.  We have also checked that the
levels of the sine-Gordon model that contribute to the minimal partition
coincide with predictions from the massless TBA for minimal models. We have
presented a detailed study of the case $n=0$ because supersymmetry allowed us
to do so. This case is certainly a little marginal, since for instance the
associated $c=0$ ``minimal model'' has no degrees of freedom and a partition
function equal to one, and so is clearly ``infinitely far'' from the
non-minimal one whose partition function grows exponentially.  This case
presents strong evidence for genuine singularities in the flow whose physical
interpretation is rather clear from polymer point of view. The natural
questions to ask now are

1) are there really singularities in the flow for $n=0$?

2) if there are indeed singularities for $n=0$, is it a marginal case or do
all $0\leq n$ values exhibit the same behavior?

Certainly for $0<n<2$ the situation is somewhat different than for $n=0$ since
the free energies of minimal and sine-Gordon models are then the same (level
crossings occur in some sectors, but only in finite numbers).

3) is there an exact scattering theory for massless flow within sine-Gordon
and between successive minimal models, at least for some values of $0\leq n$?

The study of these questions and partial answers are presented in the next
paper.

\bigskip\bigskip\centerline{{\bf Acknowledgments}}\nobreak
We would like to thank K. Intriligator, G. Moore, C. Vafa and A.B.
Zamolodchikov for useful conversations. H.S.\ was supported by the Packard
Foundation and DOE grant DE-AC-76ERO3075, while P.F.\ was supported by DOE
grant DEAC02-89ER-40509.

\appendix{A}{Polymer results}

We summarize here the polymer results obtained in this paper for the reader
with this specific interest only. Consider the total number of configurations
for a set of $L$ polymer lines of total number of edges ${\cal N}$ crossing
the cylinder of radius $R^l$ (which we denote simply by $R$ here). One has
\eqn\conf{\Omega_L^{(R)}\propto\left(\mu_L^{(R)}\right)^
{{\cal N}},\ {\cal N}>>1.}
The connectivity constant $\mu_L^{(R)}$ is such that for every $L$,
$\mu_L^{(R)}\rightarrow\mu$ as $R\rightarrow\infty$ where $\mu$ is the usual
lattice-dependent ``bulk'' connectivity constant \ref\gut{A.J. Guttman,
J.Phys. A17 (1984) 455, and references therein.}. Using finite-size scaling one
has $\mu_L^{(R)}-\mu\propto R^{-1/\nu}$, where $\nu=3/4$. Define the numbers
(where $\beta_c=1/\mu)$
\eqn\rLapp{\hbox{lim}
_{R\rightarrow\infty}(1/\mu_L^{(R)}-\beta_c)^{\nu}R^l=r^l_L.}
These numbers are universal up to a lattice-dependent overall scale (that is
$r_L/r_{L'}$ is universal). We have found that the particular numbers
$r_{2+4n}$ are in the same ratio as the (simple) $n^{\hbox{th}}$ poles of the
function $Q$ given by \Q; $Q$ is the derivative of a solution of the
Painlev\'e III equation \PIII, with boundary condition \asym.  The locations
of these poles do not have simple expressions. The first few numerical values
are

\medskip
\settabs 6\columns
\+&$1^{st}$&$2^{nd}$&$3^{rd}$&$4^{th}$&$5^{th}$\cr
\smallskip
\+&2.95708396&8.61&14.1&19.5&24.8\cr
\medskip
\noindent
We have also found that the distance $r_{2+4n}-r_{-2+4n}$ behaves as $Cr_{4n}$
for $n$ large, where $C$ is a constant close to $C\simeq 5$.

That $Q$ should have a singularity can also be established on physical grounds
using the polymer model. Indeed recall that in \FS\ we showed that
\eqn\lapl{Q(r)=\int_0^\infty e^{-(r)^{4/3}x}f(x)dx.}
where the number of configurations of a single non-contractible
loop of length ${\cal N}$ on the cylinder reads in the scaling limit
\eqn\loopconf{\omega_{R}= TR^{-7/3}\mu^{{\cal N}}sf({{\cal N}s\over R^{4/3}}).}
and we have set
\eqn\scal{r=\left({\beta_c-\beta\over s\beta_c}\right)^{3/4}R.}
while $s$ is some lattice-dependent rescaling factor.  The dominant behavior
of the function $f$ follows from the loss of entropy due to confinement that
is $f(y)\simeq y^{\alpha}\hbox{exp}(-cy)$ where $\alpha,c$ are constants to be
determined. Plugging this result back into the Laplace transform \lapl\ we see
immediately that $c=(r_2)^{4/3}$ and $\alpha=0$. That the connectivity
constant for a non-contractible loop and for two propagating legs ($L=2$)
should be the same appears reasonable, and it is proven by the quantum-group
arguments in the text.

Some of these results could likely be extended to the other number of legs, in
particular odd ones. However the connection with the PIII equation, which has
been our main analytical tool here, is possible only for a number of legs equal
to two modulo four, which is characteristic of the Ramond sector in the $N$=2
point of view.

\vfill\eject
\listrefs
\centerline{\bf Figure Captions}
\bigskip

Figure 1: Data for the running central charge $c(mR)$ for $n=\sqrt{2}$ obtained
from the lattice model and analytic continuation of massive TBA

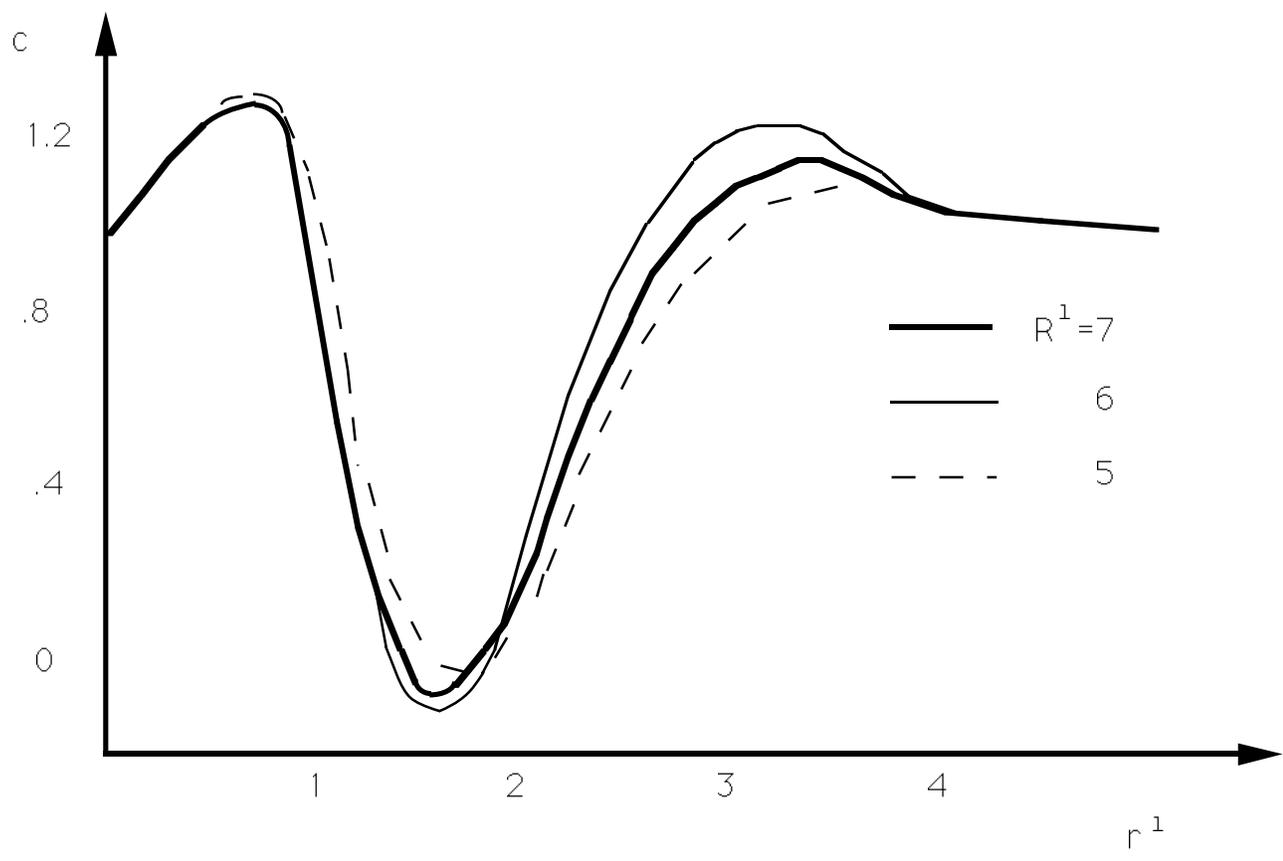
Figure 2: Data for the running central charge at $n=-1$ obtained from the
lattice model. Observe the appearance of additional extrema in the IR as
$R^l$ increases.

Figure 3: The locus of zeroes for the Ising model partition function.
Perturbation by a purely imaginary mass moves along the circles.

Figure 4a: A set of $L$ polymer lines propagating on the cylinder

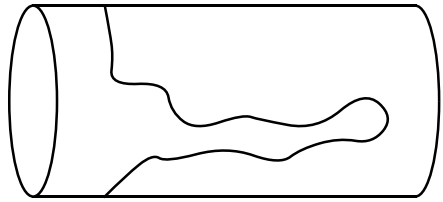
Figure 4b: Growth of an infinite arm at $r_2$

Figure 5: Discontinuity of the running subtracted central charge in the Ramond
sector ($n=0$) at $r_2$

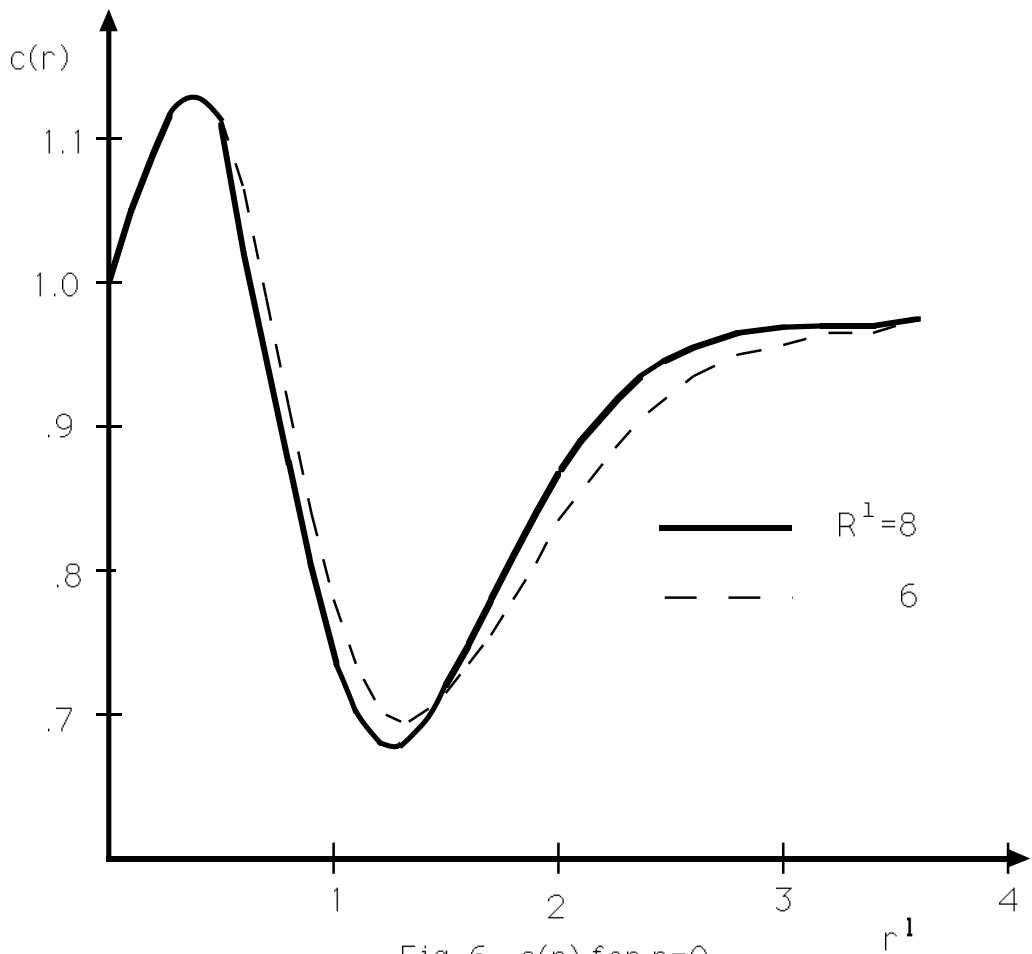
Figure 6: Running central charge $c(mR)$ for $n=0$ obtained from the lattice

Figure 7: Second derivative of $c(r=mR)$ with respect to $r$ for $n=0$.

\bye